%% file: main.tex
\documentclass[aps,prl,reprint,twocolumn,groupedaddress]{revtex4-1}

\usepackage{amsmath, amssymb, natbib, graphicx, url, cleveref}
\Crefformat{figure}{#2Fig.~#1#3}

\usepackage{xcolor, subfigure}
\newcommand*{\br}{\boldsymbol{r}}

\newcommand\add[1]{\textcolor{black}{#1}}
\newcommand{\replace}[2]{\textcolor{black}{#2}}

\usepackage{chngcntr}
\counterwithout{paragraph}{subsubsection}

\begin{document}


\title{Finding the Differences: Classical Nucleation Perspective on Homogeneous Melting and Freezing of Hard Spheres}


\author{Willem Gispen}
\email[]{w.h.gispen2@uu.nl}

\author{Marjolein Dijkstra}
\email[]{m.dijkstra@uu.nl}

\affiliation{Soft Condensed Matter and Biophysics, Debye Institute for Nanomaterials Science, Utrecht University}


\date{\today}

\begin{abstract}

By employing  brute-force molecular dynamics, umbrella sampling, and seeding simulations, we investigate homogeneous nucleation during melting and freezing of hard spheres. We provide insights into these opposing phase transitions from the standpoint of classical nucleation theory. We observe that melting has both a lower driving force and a lower interfacial tension than freezing. The lower driving force arises from the vicinity of a spinodal instability in the solid and from a strain energy. The lower interfacial tension implies that the Tolman lengths associated with melting and freezing have opposite signs, a phenomenon that we interpret with Turnbull's rule. Despite these asymmetries, the nucleation rates for freezing and melting are found to be comparable. 

\end{abstract}







\maketitle


Melting and freezing are opposing pathways by which matter transitions between solid and liquid phases. These processes have significant implications across various scientific disciplines, including materials science, geophysics, biology, and atmospheric sciences. Understanding them is crucial in numerous practical applications. For instance, in metallurgy, knowledge of freezing and melting is essential for processes like casting, welding, and shaping. In the pharmaceutical industry, crystallization and melting play a crucial role in influencing the solubility and formulation of drugs.

Both melting and freezing  most commonly occur through heterogeneous nucleation on external surfaces or internal defects. Nonetheless, homogeneous nucleation remains a useful starting point for understanding the thermodynamics of these first-order phase transitions. For example, knowledge of the homogeneous freezing rate can be used to predict the heterogeneous freezing rate~\cite{yuan_rseeds_2023}. Experimentally, the involvement of surfaces and defects makes the study of  homogeneous melting and freezing challenging. However, it is possible to mitigate these effects, for instance,  external surfaces can be coated~\cite{daeges_superheating_1986},  and defects can be annealed  using heating cycles~\cite{wang_direct_2015}. This has enabled the experimental observation of homogeneous melting in a system of thermally responsive microgel colloids~\cite{alsayed_premelting_2005,wang_imaging_2012,wang_direct_2015}.

The key factor in homogeneous nucleation is the nucleation rate. Classical nucleation theory (CNT) plays a paramount role  for  qualitatively  understanding  this  rate. 
CNT consists of three components, the driving force, interfacial tension, and kinetic prefactor.  Notably, the kinetic prefactor and  driving force can be accurately determined using the bulk equations of state. 
This leaves the interfacial tension as the primary unknown in CNT. 
Phenomenological rules exist for predicting trends in the interfacial tension. For instance, Turnbull's phenomenological rule \cite{turnbull_formation_2004} asserts that the interfacial tension is proportional to the melting enthalpy. Additionally, Tolman \cite{tolman_effect_1949} argued that it varies with the curvature of the interface. 

In the past two decades, it has become possible to test these predictions using computer simulations. These simulations provide a unique tool for uncovering the mechanisms and thermodynamics of nucleation. 
For instance, \citet{sanchez-burgos_equivalence_2020} demonstrated that the Tolman length is equivalent for condensation and cavitation. However, much less is known about the interfacial tension governing the melting transition and whether a similar symmetry exists  between freezing and melting. One reason for this knowledge gap is that melting has traditionally been considered a one-sided instability of the solid phase, with most research focusing on the role of various types of defects that lead to catastrophic melting at the limit of superheating~\cite{wang_homogeneous_2018,charaborty_relationship_2009,forsblom_how_2005,gomez_dislocation_2003,jin_melting_2001,samanta_microscopic_2014}.  Due to the emphasis on the superheat limit, numerous unresolved questions persist regarding the mechanism and thermodynamics of melting {prior} to reaching the superheat limit. Furthermore, the focus on defects has cast significant doubts on the suitability of  CNT  for effectively describing the melting process~\cite{charaborty_relationship_2009,wang_homogeneous_2018,samanta_microscopic_2014}. \add{For example, it has never been tested whether CNT can succesfully predict the melting rate.}

In this paper, we assess the applicability of CNT in  describing  melting and freezing of hard spheres. Using computer simulations, we investigate homogeneous melting and freezing within the solid-fluid coexistence region. Our findings reveal several asymmetries between melting and freezing, offering a unique perspective on the predictions of Turnbull and Tolman.

We first perform molecular dynamics simulations to elucidate  the mechanism of homogeneous melting. We prepare a surface-free and defect-free face-centered cubic (fcc) crystal consisting of $N=2\times 10^4$ nearly hard spheres that interact with a Weeks-Chandler-Andersen (WCA) potential. This potential has been extensively used in previous research for modeling hard spheres via  molecular dynamics. Specifically, prior studies~\cite{filion_simulation_2011,gispen_brute-force_2023}  have shown that this potential effectively maps freezing rates to hard spheres when defining  an effective hard-sphere diameter $\sigma_{\mathrm{eff}}$, such that the freezing density~\cite{filion_simulation_2011} aligns with that of hard spheres~\cite{frenkel2001understanding}. 

\begin{figure}[]
    \centering
    \includegraphics[width=\linewidth]{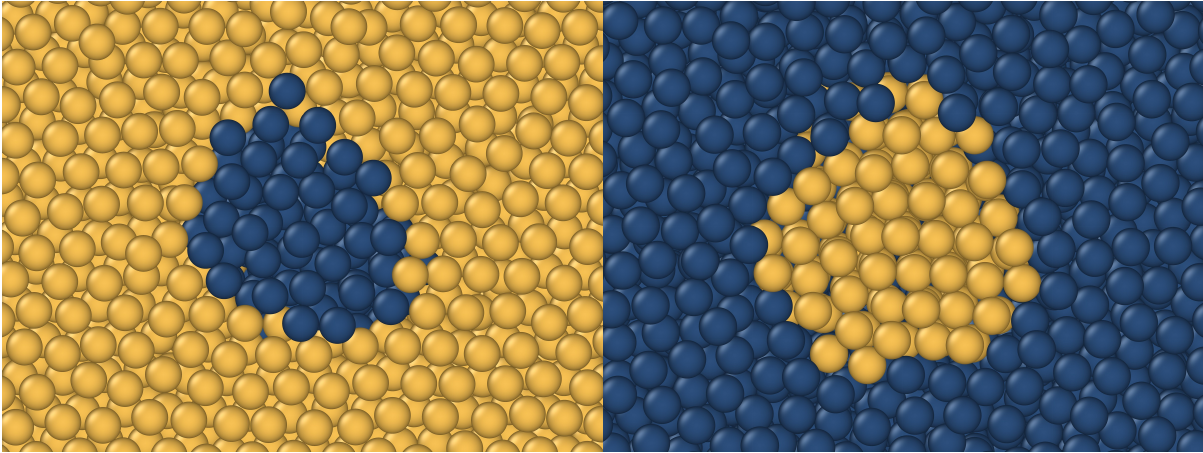}
    \caption{
    Cut-through images of nucleation during homogeneous melting (left) and freezing (right) of hard spheres. The nuclei are identified using the Steinhardt bond order parameter $\bar{q}_6$. Yellow corresponds to solid-like particles, dark blue to fluid-like particles.
    }
    \label{fig:spontaneous}
\end{figure}

In the following, we use this effective hard-sphere diameter to calculate the effective packing fraction $\eta=N \pi \sigma_{\mathrm{eff}}^3/ 6 V$. \add{We perform all our simulations in the coexistence region between the freezing point $\eta_{\mathrm{fr}}=0.492$ and melting point $\eta_{\mathrm{m}}=0.544$} \cite{frenkel2001understanding}. Firstly, we perform brute-force canonical ensemble ($NVT$) simulations of the fcc crystal from $\eta=0.503$ to $0.508$. 
In \Cref{fig:spontaneous}, we show a fluid nucleus at $\eta=0.508$. \add{For comparison, we also show} a crystal nucleus \add{from previous simulations \cite{gispen_brute-force_2023} at $\eta=0.528$}. We differentiate  between fluid-like (dark blue) and solid-like (yellow) particles using the averaged Steinhardt bond order parameter $\bar{q}_6$~\cite{lechner_accurate_2008}.  Employing the mislabeling scheme~\cite{espinosa_seeding_2016}, we calculate a pressure-dependent threshold $\bar{q}_6^*(P)$ and classify particles as fluid-like when $\bar{q}_6 < \bar{q}_6^*(P)$ and solid-like when $\bar{q}_6 > \bar{q}_6^*(P)$. For additional details, please refer to the Supplemental Materials (SM)~\cite{SM}. 
 Similar to freezing, the fluid nucleus depicted in \Cref{fig:spontaneous} has an approximately spherical shape. We did not observe nuclei with a lentil or oblate shape\add{, not even in the region close to the melting point where they are theoretically predicted \cite{brener_elastic_1999,motorin_kinetics_1984}. Please see the SM ~\cite{SM} for additional illustrations of the shapes of the fluid nuclei.} Therefore, we assume a spherical shape in our subsequent analysis. The formation of a nucleus can take a long time, but once it reaches a critical size, it rapidly expands, leading to  melting of the entire system. This observation, that melting follows a nucleation and growth scenario, aligns with prior findings \cite{wang_homogeneous_2018} and reinforces the applicability of CNT.

\begin{figure}[!t]
    \centering
    \includegraphics[width=\linewidth]{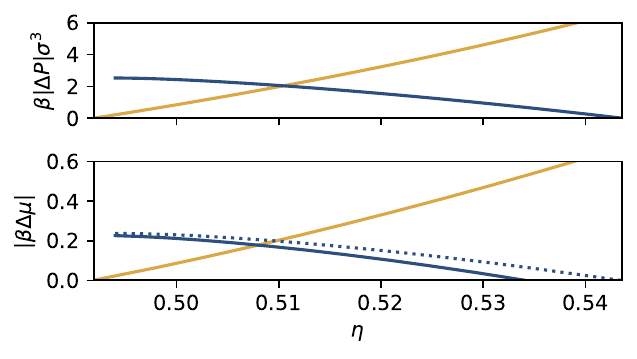}
    \caption{(Top) Pressure difference from coexistence $\beta |P - P_{\mathrm{coex}}| \sigma^3$ for the fluid (yellow) and solid (dark blue) phase of hard spheres, as a function of packing fraction $\eta$. (Bottom) Driving force for melting and freezing  as a function of packing fraction $\eta$. The yellow (dotted blue) line is the difference in chemical potential between the solid and fluid phase as a function of the packing fraction of the fluid (solid) phase. 
    The solid blue line is the effective driving force $|\Delta\mu_{\mathrm{eff}}|=|\Delta\mu|-E_{\mathrm{strain}}$ for melting.} 
    \label{fig:driving-force}
\end{figure}

\begin{figure}[!ht]
  \includegraphics[width=\linewidth]{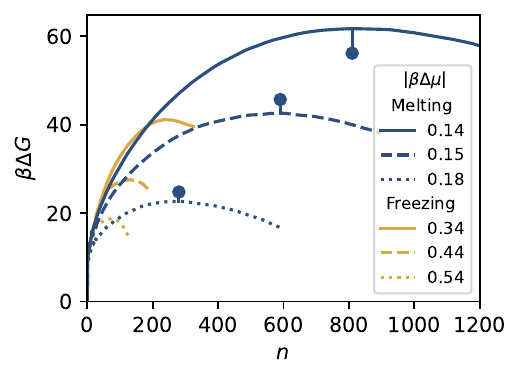}
  \caption{Gibbs free energy $\beta \Delta G$ as a function of nucleus size $n$ for melting and freezing as obtained from umbrella sampling, for three different driving forces $\beta\Delta\mu$. The freezing barriers (yellow) are taken from \citet{auer_prediction_2001}. The melting barriers (blue) are determined in this work. 
  For melting, the blue dots are the CNT approximations  of the barrier height $\Delta G^*=n^* |\Delta\mu_{\mathrm{eff}}|/2$, with $n^*$ the critical nucleus size.
  }
  \label{fig:barriers}
\end{figure}

\begin{figure*}[!htbp]
  \includegraphics[width=\textwidth]{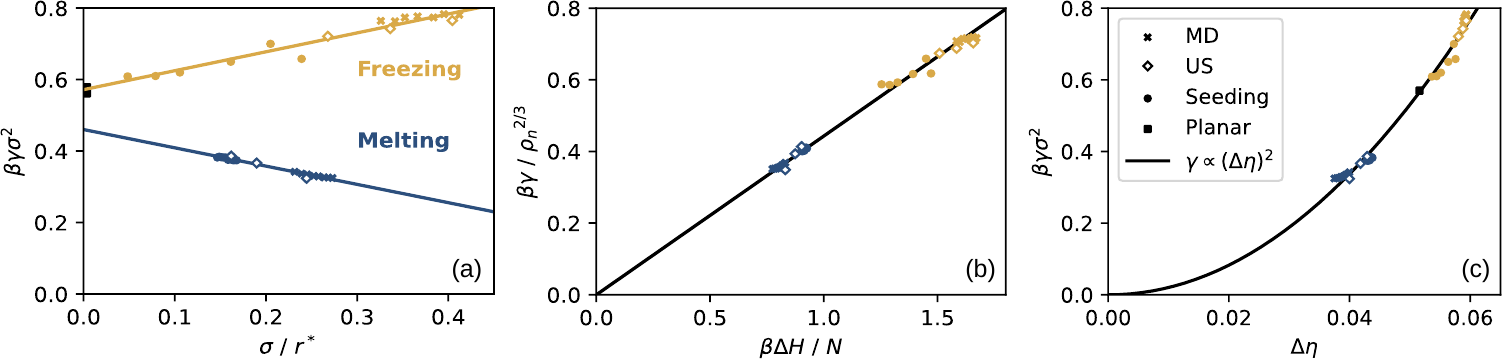}
  \caption{Interfacial tension $\gamma$ as a function of (a) inverse critical radius $\sigma / r^*$, (b) enthalpy difference $\beta\Delta H / N$ between fluid and solid, and (c) difference in packing fraction $\eta$ between fluid and solid. The interfacial tension is calculated using CNT from molecular dynamics (MD, crosses), umbrella sampling (US, diamonds), and seeding (circles). \add{The lines are either linear (a,b) or power law (c) fits of the data.} For freezing, we used data from \citet{gispen_brute-force_2023} (MD), \citet{espinosa_seeding_2016} (seeding) and \citet{filion_crystal_2010} (US). Following \citet{turnbull_formation_2004}, we normalize $\gamma$ with the density $\rho_n$ of the nucleating phase for figure (b).}
  \label{fig:gamma-R}
\end{figure*}

To quantitatively describe the thermodynamics of melting and freezing using CNT, we first calculate the driving force for nucleation, which is determined by the difference in chemical potential $|\Delta\mu|$ between the fluid and solid phase.  We calculate this difference by employing thermodynamic integration of empirical equations of state~\cite{liu_carnahan-starling_2021,wang_homogeneous_2018} from coexistence~\cite{frenkel2001understanding}. In \Cref{fig:driving-force}, we present the driving force $|\Delta\mu|$ for both the melting and freezing of hard spheres as a function of packing fraction $\eta$. Surprisingly, the driving force for melting is significantly lower than that for freezing. According to the Gibbs-Duhem equation $\partial |\Delta\mu| / \partial P=\Delta (1/\rho)$, the derivative \replace{of the driving force with respect to pressure}{$\partial |\Delta\mu| / \partial P$} should be nearly equal for melting and freezing. However, within the solid-fluid coexistence region, the pressure of the solid changes much less than that of the fluid. To illustrate this, we also plot the pressure difference from coexistence $\beta |P - P_{\mathrm{coex}}| \sigma^3$ in \Cref{fig:driving-force}. 
The reduced pressure variation  for the solid phase is associated with the spinodal instability occuring at $\eta=0.494$~\cite{wang_homogeneous_2018}, where, by definition, the derivative $\partial P / \partial \eta$ becomes zero. In summary, the primary reason for the lower driving force for melting is the presence  of a spinodal instability.
However, the driving force for melting is also reduced by  
strain energy~\cite{motorin_kinetics_1984,jin_melting_1998,lu_homogeneous_1998,bai_nature_2005}. The reduced density within a fluid nucleus compels the surrounding crystal to deform, incurring an additional free-energy cost that is proportional to the volume of the fluid nucleus
\begin{equation} \label{eq:cnt}
    \Delta G=\gamma 4\pi r^2  -|\Delta \mu| \frac{4\pi}{3} r^3 \rho_f + E_{\mathrm{strain}} \frac{4\pi}{3} r^3.
\end{equation}
Here, $\Delta G$ denotes the Gibbs free energy  for nucleation, $\gamma$ is the solid-fluid interfacial tension, $r$ and $\rho_f$ are the radius and density of the fluid nucleus, respectively. Additionally, $E_{\mathrm{strain}}$ corresponds to the strain energy given by 
\begin{equation}
     E_{\mathrm{strain}}=\frac{2 C_s B_f}{4 C_s + 3 B_f} \left ( \frac{1/\eta_f - 1/\eta_s}{1/\eta_s}\right )^2.
\end{equation}
In this expression, $C_s$ represents the shear modulus of the solid~\cite{wang_homogeneous_2018}, $B_f$ the bulk modulus of the fluid~\cite{liu_carnahan-starling_2021}, and the term within brackets denotes the relative volume change due to melting. The strain energy lowers the effective driving force $|\Delta\mu_{\mathrm{eff}}|=|\Delta\mu| - E_{\mathrm{strain}}/\rho_f$. Given the very low driving force for melting, it is surprising that we have observed spontaneous melting at $\eta_{\mathrm{eff}}=0.508$, where the effective driving force is $0.18 ~k_B T$. In contrast, spontaneous freezing can only be observed at $\eta \geq 0.528$~\cite{gispen_brute-force_2023}, where the driving force is $\geq 0.44 ~k_B T$.

To understand this surprising asymmetry, we separate thermodynamics from kinetics.\replace{We compute the free-energy barrier for homogeneous melting using the umbrella sampling technique~\cite{auer_prediction_2001}. We once again simulate $N=2\times 10^4$ particles interacting with a WCA potential, but this time in the isobaric-isothermal ($NPT$) ensemble.}{} Starting from a perfect fcc lattice, we measure the Gibbs free-energy barrier $\beta \Delta G (n)$ required to form a fluid nucleus of size $n$ \replace{}{using the umbrella sampling technique~\cite{auer_prediction_2001}}. In \Cref{fig:barriers}, we present these free-energy barriers for three different pressures, corresponding to effective driving forces  $|\beta\Delta\mu_{\mathrm{eff}}|=0.14, 0.15,$ and $0.18$. According to CNT, the barrier height $\Delta G^*$ and critical nucleus size $n^*$ are related as $\Delta G^*=n^* |\Delta\mu_{\mathrm{eff}}|/2$. We plot this CNT approximation as blue dots in \Cref{fig:barriers}, showing that it is a reasonable approximation, with a maximum error of $5 k_B T$ at the lowest supersaturation. In the same figure, we also display the freezing barriers computed by \citet{auer_prediction_2001}. When we compare their freezing barriers with our melting barriers, we observe that the critical nucleus size for melting is much larger than for freezing. For instance, the freezing barrier at \replace{a driving force of}{} $|\beta\Delta\mu|=0.34$ and the melting barrier at $|\beta\Delta\mu_{\mathrm{eff}}|=0.15$ both have a height of approximately $42~k_B T$, but the critical nucleus size is approximately $590$ for melting and $230$ for freezing.
The same comparison reveals that melting has a significantly lower effective driving force for a nucleation barrier of comparable  height. 

From a CNT perspective, this implies that the interfacial tension for melting must be lower than for freezing. We calculate the interfacial tension using CNT based on  brute-force~\cite{gispen_brute-force_2023},  seeding~\cite{espinosa_seeding_2016}, and umbrella sampling~\cite{filion_crystal_2010} simulations,  see SM~\cite{SM}. In \Cref{fig:gamma-R}a, we plot the interfacial tension $\gamma$ as a function of the inverse critical radius $1/r^*$ for both freezing and melting. 
 \replace{Additionally, we show the planar limit $\beta\gamma\sigma^2=0.57$~\cite{royall_colloidal_2023} with a black square in \Cref{fig:gamma-R}.}{} \replace{T}{We see that t}he interfacial tension for freezing converges to the planar limit \replace{}{$\beta\gamma\sigma^2=0.57$~\cite{royall_colloidal_2023}}  as $1/r^* \to 0$\replace{. In contrast, the interfacial tension for melting}{, whereas it} extrapolates to a lower value for melting, which may be associated with the strain energy\replace{ involved in melting}. \replace{The interfacial tension increases with $1/r^*$ for freezing~\cite{montero_de_hijes_interfacial_2019,montero_de_hijes_interfacial_2020} whereas it {decreases}  for melting. Hence,}{Clearly,} the common assumption of a constant interfacial tension ~\cite{lu_homogeneous_1998,luo_nonequilibrium_2004,rethfeld_ultrafast_2002,wang_homogeneous_2018} is not suitable for hard spheres\replace{.}{, since the interfacial tension increases with $1/r^*$ for freezing~\cite{montero_de_hijes_interfacial_2019,montero_de_hijes_interfacial_2020} whereas it {decreases}  for melting.}
\replace{Our results for the interfacial tension for melting deviate from the experimental value of $\beta\gamma\sigma^2=0.84$ estimated from homogeneous nucleation~\cite{wang_imaging_2012}, but align well with the experimental value of $\beta\gamma\sigma^2=0.42$ estimated from nucleus growth~\cite{wang_direct_2015}.Qualitatively, our observation of a decreasing interfacial tension for melting is consistent with previous simulations of a Lennard-Jones system~\cite{baidakov_kinetics_2018,bai_nature_2005}. } Based on \replace{the best linear fits of the interfacial tension $\beta\gamma\sigma^2=0.57 (1 + 0.92\sigma/r^*)$ for freezing and $0.46 (1 - 1.11 \sigma/r^*)$ for melting}{linear fits of the interfacial tension}, we can estimate the Tolman lengths as $-0.46\sigma$ for freezing and $0.56\sigma$ for melting. \replace{We note that the Tolman length for melting has the same sign but a higher magnitude compared to the value $0.34\sigma$ found in experiments~\cite{wang_homogeneous_2018}.} Interestingly, we observe that the Tolman length has opposite signs but similar magnitudes for freezing and melting. In the context of boiling and condensation, opposite signs of the Tolman length have been attributed to the transition from a convex to a concave surface, i.e. a shift from positive to negative curvature~\cite{binder_overview_2016}.

\begin{figure}[!t]
    \centering
    \includegraphics[width=\linewidth]{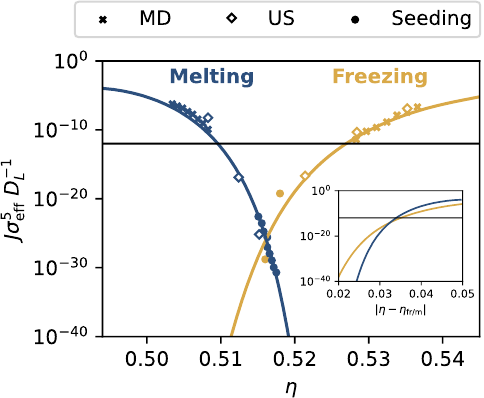}
    \caption{Nucleation rate $J\sigma_{\mathrm{eff}}^5/D_L$ of freezing and melting as a function of packing fraction $\eta$, where $D_L$ denotes the long-time diffusion coefficient. We show brute-force molecular dynamics (MD, crosses), umbrella sampling (US, diamonds), and seeding (circles \add{and lines}) predictions for the nucleation rate. The results for melting are from this work, the results for freezing are taken  from \citet{espinosa_seeding_2016} (seeding), \citet{filion_crystal_2010} (US), and \citet{gispen_brute-force_2023} (MD). The horizontal solid line is an approximate lower limit $J\sigma_{\mathrm{eff}}^5/D_L=10^{-12}$ to what is accessible to experiments on colloidal hard spheres~\cite{royall_colloidal_2023}. The inset shows the nucleation rate as a function of packing fraction difference from the freezing point $\eta_{\mathrm{fr}}$ and melting point $\eta_{\mathrm{m}}$, respectively.}
    \label{fig:rate}
\end{figure}

To understand the origin of this asymmetry in the interfacial tension, we invoke Turnbull's phenomenological rule~\cite{turnbull_formation_2004}, which states that the interfacial tension is proportional to the difference in enthalpy between the solid and fluid phases. In \Cref{fig:gamma-R}b, we also show the interfacial tension $\gamma$ as a function of this enthalpy difference $\Delta H$. Following Turnbull, we normalize $\gamma$ with the density $\rho_n$ of the nucleating phase to the power $2/3$. We see that $\gamma$ is indeed approximately proportional to $\Delta H$.
Alternatively, we can understand the lower interfacial tension by examining how it scales with the difference in packing fraction $\Delta \eta$ between the solid and fluid phases. In \Cref{fig:gamma-R}c, we plot this relationship. The best power-law fit yields an exponent of $2.02$, which reasonably describes the interfacial tension.
The power-law scaling \replace{of the interfacial tension with the density difference}{} is reminiscent of the scaling laws for surface tension near a gas-liquid critical point~\cite{rowlinson_molecular_1989}. However, it is important to note that both Turnbull's rule and the scaling laws near a critical point are not specifically  intended to describe the variation of fluid-solid interfacial tension with supersaturation. Nevertheless, they provide some insight into why the interfacial tension is lower for melting: the differences in density and enthalpy quantify the extent to which the density and structure of the two phases must change across the fluid-solid interface. Because these differences are much smaller for melting, it explains why the interfacial tension is also lower. 

So far, we have interpreted our brute-force and umbrella sampling simulations using CNT, and have shown there are significant asymmetries between freezing and melting concerning driving force and interfacial tension. However, how does this translate to the most important quantity, which is also experimentally accessible --the nucleation rate? Can CNT be employed  to \emph{predict} the nucleation rate as well? To address  these questions, we use fluid nuclei that were equilibrated with  umbrella sampling simulations as initial configurations for seeding simulations~\cite{espinosa_seeding_2016}. For eight different seeds, ranging in size from $n=800$ to $1200$, we determine the driving force necessary to make these seeds critical. In this way, we obtain the critical nucleus size $n^*$ as a function of the driving force. Subsequently, the CNT prediction for the nucleation rate is expressed as $J=J_0 \exp (-\Delta G^* / kT)$, where the barrier height is calculated as $\Delta G^*=n^* |\Delta \mu_{\mathrm{eff}}|/2$, and $J_0$ denotes the kinetic prefactor. From the seeding simulations, we determine that $J_0 \approx 50 D_L$, where $D_L$ is the long-time diffusion coefficient of the fluid phase. This kinetic prefactor is very similar to the kinetic prefactor for freezing, and  we assume it to remain constant within the fluid-solid coexistence region. \add{Using this knowledge, we can now make CNT predictions for the nucleation rate within the full fluid-solid coexistence region, please see the SM~\cite{SM} for additional details.}

In \Cref{fig:rate}, we plot the nucleation rates for freezing and melting, as evaluated using brute-force molecular dynamics (MD), umbrella sampling (US), and seeding. \add{The CNT predictions are shown with solid lines.} For freezing, we show previously published results~\cite{gispen_brute-force_2023,filion_simulation_2011,espinosa_seeding_2016}, while the melting rates are all original. We find that the CNT prediction of the melting rate via seeding agrees very well with our brute-force and umbrella sampling simulations. This good agreement was previously shown for freezing as well~\cite{espinosa_seeding_2016,gispen_brute-force_2023}, which can be considered surprising because CNT has been argued to provide highly inaccurate nucleation rate predictions, with discrepancies reaching as high as ten orders of magnitude. 

These results indicate that when accounting for the dependence of interfacial tension on curvature and pressure, CNT can offer excellent predictions using seeding for both melting and freezing of hard spheres. In the inset of \Cref{fig:rate}, we compare the nucleation rates at equal packing fraction differences from the freezing point $\eta_{\mathrm{fr}}=0.492$ and melting point $\eta_{\mathrm{m}}=0.544$, respectively~\cite{frenkel2001understanding}. Despite the differences in driving force and interfacial tension, the nucleation rate exhibits a fairly symmetric behavior. Particularly within the regime $|\eta-\eta_{\mathrm{f/m}}| > 0.034$, the freezing and melting rates are very comparable, with a difference of less than two orders of magnitude. Conversely, within the regime $|\eta-\eta_{\mathrm{f/m}}| < 0.034$, the nucleation rate $J\sigma_{\mathrm{eff}}^5/D_L$ for both melting and freezing falls below $10^{-12}$. This value serves as an approximate limit for what can be experimentally observed with colloidal hard spheres~\cite{royall_colloidal_2023}. Consequently, this region $|\eta-\eta_{\mathrm{f/m}}| < 0.034$ or equivalently $0.51 < \eta < 0.526$, represents a `forbidden zone' for homogeneous nucleation of colloidal hard spheres. Our results suggest that when phase transitions are observed, whether it is melting or freezing, homogeneous nucleation is not the driving mechanism in this regime.

In conclusion, we have studied homogeneous melting and freezing of hard spheres, with particular focus on the thermodynamic factors employed  in classical nucleation theory. We have identified several reasons for the asymmetry between melting and freezing, such as the vicinity of a  spinodal instability in the solid phase and the strain energy associated with melting. Moreover, melting exhibits both a lower driving force and a lower interfacial tension in comparison to freezing. This interfacial tension asymmetry is characterized by similar-magnitude but opposite-sign Tolman lengths for freezing and melting. We have interpreted this asymmetry through Turnbull's rule and a scaling law. Remarkably, the asymmetries in driving force and interfacial tension roughly offset each other.  When comparing the nucleation rates of freezing and melting  at equal packing fraction differences from the freezing and melting lines, respectively, we find that both processes exhibit  roughly similar nucleation rates. \add{Finally, our discovery that classical nucleation theory, augmented with an elastic strain energy correction, accurately predicts the homogeneous melting rate, holds promise for transferring it to the study of melting phenomena in atomic, molecular, and colloidal systems.}

\begin{acknowledgments}
 W.G. thanks Eduardo Sanz for useful discussions. M.D. and W.G. acknowledge funding from the European Research Council (ERC) under the European Union’s Horizon 2020 research and innovation programme (Grant agreement No. ERC-2019-ADG 884902 SoftML).
\end{acknowledgments}

\include{supplementary}

\end{document}

%% file: supplementary.tex
\onecolumngrid
\section*{Supplemental Material of ``Finding the Differences: Classical Nucleation Perspective on Homogeneous Melting and Freezing of Hard Spheres"}


\affiliation{Soft Condensed Matter and Biophysics, Debye Institute for Nanomaterials Science, Utrecht University}


\subsection{Code}
The code to generate the results in this paper are available as a zip archive in the supplemental material.

\subsection{Model system} 
\label{sec:model}
We model $N=2\times 10^4$ hard spheres using a Weeks-Chandler-Andersen (WCA) potential. The pair interaction is given by
\begin{equation*}
u \left(r_{ij}\right) = 
\begin{cases}
    4 \epsilon \left\lbrack \left(\frac{\sigma}{r_{ij}} \right)^{12} \hspace{-2mm}- \left(\frac{\sigma}{r_{ij}} \right)^6 \hspace{-2mm}+ \frac{1}{4} \right\rbrack  &r_{ij}<2^{\frac{1}{6}}\sigma\\
  0 &r_{ij}\geq2^{\frac{1}{6}}\sigma ,
\end{cases}
\label{wca}
\end{equation*}
where $r_{ij}$ is the distance between particles $i$ and $j$, $\epsilon$ is the interaction strength, and $\sigma$ is the diameter of a particle. The WCA potential is purely repulsive, and the steepness is determined by the temperature $k_BT / \epsilon$. We use $k_BT / \epsilon = 0.025$, which has been used in the past~\cite{filion_simulation_2011} to model hard spheres. At this temperature, the freezing density of the WCA system is $\rho^{\mathrm{WCA}}\sigma^3 = 0.712$~\cite{filion_simulation_2011}, whereas the freezing packing fraction of hard spheres is $\rho^{\mathrm{HS}} \sigma^3=0.9391$~\cite{frenkel2001understanding}. Therefore, to map the WCA system on hard spheres, we define an effective hard sphere diameter $\sigma_{\mathrm{eff}} / \sigma = (\rho^{\mathrm{HS}} / \rho^{\mathrm{WCA}})^{1/3} \approx 1.097$. In this way, we also define an effective packing fraction $\eta = \pi \rho \sigma_{\mathrm{eff}}^3 / 6$ for the WCA system.

\subsection{Molecular dynamics}
We perform molecular dynamics simulations with the LAMMPS molecular dynamics software~\cite{plimpton_fast_1995}. We use the velocity-Verlet algorithm with a timestep $\Delta t = 0.001 \sqrt{\beta m \sigma^2}$ to integrate Newton's equations of motion. The brute-force melting simulations are carried out in the canonical (NVT) ensemble, with a Nos\'e-Hoover thermostat keeping the temperature $T$ fixed. The seeding simulations are carried out in the isobaric-isothermal ensemble, with a Nos\'e-Hoover barostat fixing the pressure. The relaxation time of the thermostat and barostat are set to $100$ and $500$ timesteps, respectively.

For the brute-force melting simulations, the initial configuration is a perfect fcc crystal with $N=2\times 10^4$ particles. The simulation is at constant volume, and as a consequence, the pressure sharply increases when the system melts. We use this increased pressure as a marker to identify a spontaneous melting event. To compute the nucleation rate $J$, we use the same procedure as we recently described in Ref. \citenum{gispen_brute-force_2023}. When a spontaneous nucleation event occurs, we immediately stop the  simulation and note the final simulation time $t_i$. When a simulation $i$ does not melt, we also note the final simulation time $t_i$. From the number of nucleation events $\ell$, the volume $V$, and the total simulation time $t=\sum t_i$, we compute the nucleation rate
\begin{equation*}
    J = \ell / (V t).
\end{equation*}
In \Cref{tab:melting-rates}, we specify the effective packing fractions and the number of simulations per packing fraction that we used to compute the nucleation rate. The nucleation rate is normalized by the long-time diffusion coefficient $D_L$ of the fluid phase at the same packing fraction. We computed $D_L$ from independent simulations of the fluid phase.

\begin{table}[]
    \centering
    \begin{tabular}{c|c|c|c|c}
   $\eta_s$ & $N \sigma^3/V$ & $\ell$ & $L$ & $J \sigma_{\mathrm{eff}}^5 D_L^{-1}$ \\
       \hline
        0.5034 & 0.729 & 32 & 32 & $3 \times 10^{ -6 }$ \\
        0.5041 & 0.730 & 32 & 32 & $2 \times 10^{ -6 }$ \\
        0.5048 & 0.731 & 32 & 32 & $8 \times 10^{ -7 }$ \\
        0.5055 & 0.732 & 32 & 32 & $3 \times 10^{ -7 }$ \\
        0.5062 & 0.733 & 32 & 32 & $1 \times 10^{ -7 }$ \\
        0.5069 & 0.734 & 22 & 32 & $2 \times 10^{ -8 }$ \\
        0.5076 & 0.735 & 22 & 32 & $5 \times 10^{ -9 }$ \\
        0.5083 & 0.736 & 8 & 128 & $7 \times 10^{ -10 }$ \\
    \end{tabular}
    \caption{Homogeneous melting rates $J \sigma_{\mathrm{eff}}^5 D_L^{-1}$ of hard spheres as a function of effective packing fraction $\eta_s=\pi N \sigma_{\mathrm{eff}}^3 / 6V$ of the solid phase. The effective packing fraction is computed from the density of the WCA system $N\sigma^3/V$ given in the second column. The nucleation rates are estimated from $\ell$ spontaneous nucleation events observed in a total of $L$ simulations.}
    \label{tab:melting-rates}
\end{table}

\subsection{Bond orientational order parameters}
To find the fluid nucleus, we use the mislabeling criterion~\cite{espinosa_seeding_2016} based on the locally averaged bond order parameter $\bar{q}_{6}$ ~\cite{steinhardt_bond-orientational_1983, lechner_accurate_2008}. To compute $\bar{q}_{6}$, we first determine the twelve nearest neighbors of particle $i$. The local density is then projected on spherical harmonics
$$
 q_{lm}(i) = \frac{1}{12} \sum_{j=1}^{12} Y_{lm}(\br_j - \br_i).
$$
These $q_{lm}(i)$'s are averaged over the twelve nearest neighbors of $i$ to obtain
$$
 \bar{q}_{lm}(i) = \frac{1}{13} \left ( q_{lm}(i) + \sum_{j=1}^{12} q_{lm}(j) \right ).
$$
Finally, the averaged bond order parameter is calculated with
$$
 \bar{q}_{l}(i) = \sqrt{\frac{4\pi}{2 l + 1}\sum_{m} |\bar{q}_{lm}(i)|^2}.
$$
Using the mislabeling criterion, we then determine the threshold $\bar{q}_6^*$ that best separates the fluid from the solid phase. This threshold is pressure-dependent, and we only determined the threshold for pressures below the solid-fluid coexistence pressure. In this regime, we fit the mislabeling threshold linearly as $\bar{q}_6^* = 0.148 + 0.012 ~\beta P \sigma_{\mathrm{eff}}^3$. This provides a local solid-fluid classification: particles with $ \bar{q}_6 > \bar{q}_6^*$ are solid-like, and particles with $\bar{q}_6 < \bar{q}_6^*$ are fluid-like. Next, two fluid-like particles $i$ and $j$ belong to the same cluster if $i$ is one of the twelve nearest neighbors of $j$, or vice versa, and the largest cluster of fluid-like particles is the fluid nucleus. We use the freud library~\cite{ramasubramani_freud_2020} to calculate the bond orientational order parameters and to find the largest cluster.

\subsection{Umbrella sampling}
We use umbrella sampling~\cite{auer_prediction_2001} to measure the free-energy barrier for homogeneous melting of a WCA system. We use the HOOMD code~\cite{anderson_hoomd-blue_2020} to perform Monte Carlo (MC) simulations in the isobaric-isothermal (NPT) ensemble. As described in Section B, the temperature is fixed at $k_BT/\epsilon=0.025$. We study the melting barrier at three different pressures $\beta P \sigma_{\mathrm{eff}}^3=9.56, 9.76, 9.89$. These pressures correspond to effective packing fractions of the solid of $\eta=0.508, 0.512, 0.515$, respectively. The pressure is kept constant using anisotropic volume moves after every MC cycle. The initial configuration for each simulation is a perfect fcc crystal of $N=2\times 10^4$ particles.

After every trajectory of $10$ MC cycles, we accept or reject the trajectory with a probability $p = \min(1, \exp(-\Delta U / k_BT)$, where $\Delta U$ is the change in the umbrella sampling potential
\begin{equation*}
    U (n) = \frac{1}{2} \lambda (n - \hat{n})^2.
\end{equation*}
Here $n$ is the size of the largest nucleus, $\hat{n}$ is the target nucleus size, and $\lambda=0.04 k_BT$ the coupling strength. After this accept/reject step, we add the largest nucleus size to the nucleus size distribution. We perform $270$ independent simulations, of which $60$ for $\beta P \sigma_{\mathrm{eff}}^3=9.56$, $90$ for $\beta P \sigma_{\mathrm{eff}}^3=9.76$, and $120$ for $\beta P \sigma_{\mathrm{eff}}^3=9.89$. Each simulation is for a different target size $\hat{n}$, which we vary in steps of $10$. We vary the target size $\hat{n}$ from $n_0 = 20$ to $600$, from $n_0 = 10$ to $900$, and from $n_0= 10$ to $1200$, respectively, for the different pressures. For each target size, we then approximate the local gradient of the free-energy barrier from the biased nucleus size distribution as
\begin{equation*}
   \frac{\partial G(\hat{n})}{\partial n} = \frac{\hat{n} - \bar{n}}{\mathrm{Var}(n)} 
\end{equation*}
where $\bar{n}$ and $\mathrm{Var}(n)$ are the mean and variance of the biased nucleus size distribution, respectively~\cite{kastner_bridging_2005}.

In addition to the simulations with the umbrella sampling potential, we also perform unbiased simulations of the metastable solid phase. In this case, we measure the full nucleus size distribution, so not only of the largest nucleus in the system.

We combine the unbiased and biased simulations to calculate the full melting barrier as follows. The initial part of the barrier is calculated from the unbiased simulation as
\begin{equation*}
    \beta G (n) = -(\log p(n) - \log p(0))
\end{equation*}
where $p(n)$ is the fraction of nuclei of size $n$, and $p(0)$ is the fraction of solid-like particles. Then, the local gradients computed from the biased simulations are integrated with a trapezoidal integration scheme. The trapezoidal integration starts at the lowest target size $n_0$ with initial value $G(n_0)$ from the unbiased simulations.

In \Cref{tab:umbrella}, we show the critical nucleus sizes and barrier heights computed with umbrella sampling in detail. As discussed in the next section F, for umbrella sampling we compute the critical nucleus size $n^*_{\mathrm{CNT}}$ and the interfacial tension from the barrier height $\beta\Delta G^*$ using classical nucleation theory. For reference, we also give the critical nucleus size $n^*$ as measured by the mislabeling criterion in \Cref{tab:umbrella}.

\begin{figure}[]
    \centering
    \includegraphics[width=0.7\linewidth]{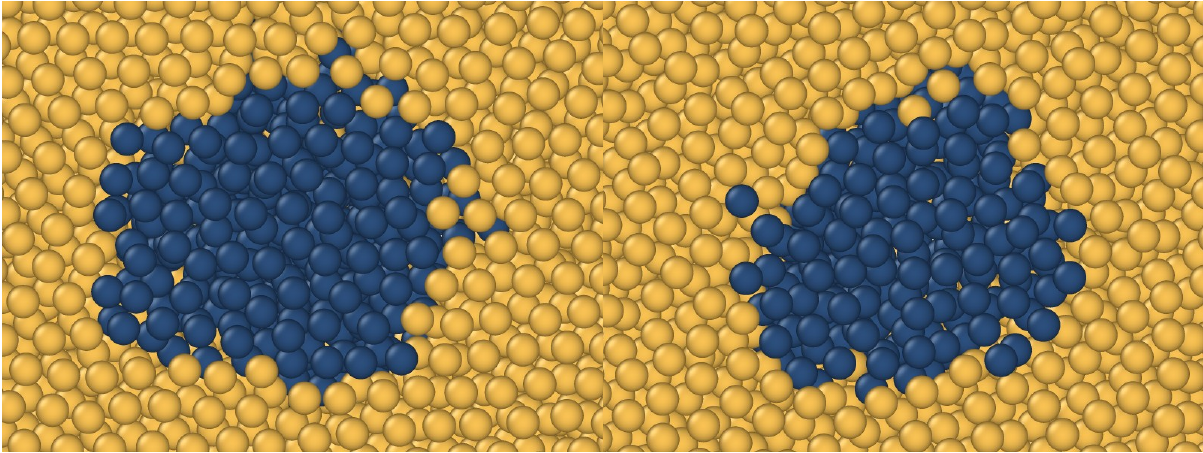}
    \caption{
    Cut-through images of nucleation during homogeneous melting of hard spheres. The nuclei are identified using the Steinhardt bond order parameter $\bar{q}_6$. Yellow corresponds to solid-like particles, dark blue to fluid-like particles. (Left) Fluid nucleus of size $n\approx 600$ at pressure $\beta P \sigma_{\mathrm{eff}}^3=9.56$. (Right)  Fluid nucleus of size $n\approx 600$ at pressure $\beta P \sigma_{\mathrm{eff}}^3=11.49$.
    }
    \label{fig:shape}
\end{figure}

\subsection{Shape of fluid nuclei}
\add{In \Cref{fig:shape}, we present fluid nuclei of size $n=600$ under two distinct conditions: one at a pressure of $\beta P \sigma_{\mathrm{eff}}^3=9.56$, where we measured the full nucleation barrier, and another at an additional pressure $\beta P \sigma_{\mathrm{eff}}^3=11.49$. The latter pressure corresponds to a packing fraction of $\eta=0.539$ for the solid. As can be seen from Fig. 2 in the main text, this is within the `forbidden zone', which means that the strain energy term surpasses $|\beta \Delta\mu|$, resulting in a negative effective driving force with absolute magnitude $|\beta\Delta\mu_{\mathrm{eff}}|=0.05$. In this regime, lentil- or oblate-shaped nuclei were predicted \cite{motorin_kinetics_1984, brener_elastic_1999}. As can be seen from \Cref{fig:shape}, these nuclei are roughly spherical instead. We have not observed lentil or oblate shaped nuclei in any of our simulations.}

\begin{table}[]
    \centering
    \begin{tabular}{c|c|c|c|c|c|c|c}
      $\beta P \sigma_{\mathrm{eff}}^3$ & $n^*$ & $n^*_{\mathrm{CNT}}$& $\eta_s$ & $\eta_{fl}$ & $|\beta\Delta\mu_{\mathrm{eff}}|$ & $\beta\gamma\sigma_{\mathrm{eff}}^2$ & $\beta \Delta G^*$ \\ \hline
    9.56 & 280 & 256 & 0.5083 & 0.4683 & 0.177 & 0.324 & 22.7 \\
    9.76 & 590 & 550 & 0.5124 & 0.4706 & 0.155 & 0.367 & 42.6 \\
    9.89 & 810 & 888 & 0.5152 & 0.4723 & 0.139 & 0.386 & 61.7 \\
    \end{tabular}
    \caption{Umbrella sampling results for the critical nucleus size $n^*$, interfacial tension $\gamma$, and barrier height $\beta\Delta G^*$ for melting of hard spheres as a function of pressure $\beta P \sigma_{\mathrm{eff}}^3$. $\eta_s$ is the effective packing fraction of the solid phase, $\eta_{fl}$ the effective packing fraction of the fluid phase at the same pressure $\beta P \sigma_{\mathrm{eff}}^3$, and $\beta\Delta\mu_{\mathrm{eff}}$ is the effective driving force for melting under those conditions.}
    \label{tab:umbrella}
\end{table}


\subsection{Interfacial tension from classical nucleation theory} 
\label{sec:cnt}
According to classical nucleation theory, the interfacial tension $\gamma$, the critical nucleus size $n^*$, the barrier height $\Delta G^*$, and the nucleation rate $J$ are related as
\begin{subequations}
    \label{eq:cnt-full}
    \begin{align}
        \label{eq:cnt-full1}
       \gamma^3 &= {3 \Delta G^* |\Delta\mu_{\mathrm{eff}}|^2 \rho_f^2 } / {16\pi}, \\
        \Delta G^* &= n^*|\Delta\mu_{\mathrm{eff}}|/2, \\
        J &= J_0 \exp (-\Delta G^* / kT),
    \end{align}
\end{subequations}
where $J_0$ is the kinetic prefactor and $\rho_f$ the density of the fluid. Note that the first two expressions follow from the classical nucleation theory expression for the barrier
\begin{equation}
    \Delta G = \gamma 4\pi r^2  -|\Delta \mu_{\mathrm{eff}}| \frac{4\pi}{3} r^3 \rho_f.
\end{equation}
Here  size $n$ and radius $r$ of the nucleus are related as $n = 4\pi r^3 \rho_f/3$. If $J_0$ and $|\Delta\mu_{\mathrm{eff}}|$ are fixed, we can use these relations to freely convert between $\gamma$, $\Delta G^*$, $n^*$, $r^*$, and $J$. In Fig. 4 of the main text, we show the interfacial tension and critical radius computed from molecular dynamics, umbrella sampling, and seeding simulations. For the molecular dynamics simulations, we computed these from the nucleation rate $J$. For the umbrella sampling simulations, we converted from the barrier height $\Delta G^*$. For the seeding simulations, we converted from the critical nucleus size $n^*$.

\subsection{Seeding}
We used the fluid nuclei equilibrated in the umbrella sampling simulations as initial configurations for seeding simulations. Normally, the initial configuration in seeding simulations is obtained by cutting a spherical nucleus from a bulk equilibrated phase and inserting it into a bulk equilibrated parent phase. In this case, that would mean inserting a spherical fluid nucleus into a bulk equilibrated fcc crystal. Unfortunately, we found that this resulted in `stuck' configurations: the nucleus size did not change significantly over time. Therefore, we used the final configurations from umbrella sampling instead.

For eight different seeds, ranging in size from $n=800$ to $1200$, we determined the critical pressure $P^*$, i.e.\ the pressure for which the seed grows and shrinks with $50\%$ probability. This was done with $NPT$ molecular dynamics simulations. From the critical pressure $P^*$, we can compute the effective driving force $\Delta\mu_{\mathrm{eff}}$. As described in Section F, from the critical nucleus size $n^*$ of the seed and the effective driving force, we can compute the interfacial tension $\gamma$. The interfacial tension is fitted linearly as a function of pressure. From this fit, we can then compute the nucleation barrier and nucleation rate using \Cref{eq:cnt-full}. 

The only missing ingredient is the kinetic prefactor $J_0$, which we calculated as follows. Following \citet{auer_prediction_2001} and \citet{espinosa_seeding_2016}, we calculate the attachment rate $f^+$ from the time evolution of the nucleus size $n(t)$ as
\begin{equation*}
    f^+ = \frac{\langle (n(t) - n^*)^2 \rangle}{2 t},
\end{equation*}
i.e.\ we fit the linear part of the nucleus size time dependence. Then the kinetic prefactor is given by\cite{espinosa_seeding_2016}
\begin{equation*}
    J_0 = \rho_s f^+ \sqrt{|\beta\Delta\mu_{\mathrm{eff}}| / (6\pi n^*)}.
\end{equation*}
We find that the kinetic prefactor depends only weakly on packing fraction, so we approximate it with the mean value $J_0 \approx 50 D_L$, where $D_L$ is the long-time diffusion coefficient of the fluid phase at the same packing fraction. We determined $D_L$ from independent simulations of the fluid phase.

In \Cref{tab:seeding}, we show the critical nucleus sizes and barrier heights computed with seeding simulations in detail. More precisely, we measure the critical nucleus size $n^*$ using seeding and compute the interfacial tension $\gamma$ and the barrier height using \Cref{eq:cnt-full}.

\begin{table}[]
    \centering
    \begin{tabular}{c|c|c|c|c|c|c}
      $\beta P \sigma_{\mathrm{eff}}^3$ & $n^*$& $\eta_s$ & $\eta_f$ & $|\beta\Delta\mu_{\mathrm{eff}}|$ & $\beta\gamma\sigma_{\mathrm{eff}}^2$ & $\beta \Delta G^*$ \\ \hline
    9.90 & 800 & 0.5151 & 0.4722 & 0.121 & 0.374 & 48.4 \\
    9.93 & 850 & 0.5155 & 0.4725 & 0.118 & 0.374 & 50.0 \\
    9.94 & 900 & 0.5158 & 0.4726 & 0.116 & 0.377 & 52.4 \\
    9.97 & 950 & 0.5163 & 0.4729 & 0.113 & 0.376 & 53.7 \\
    9.97 & 1000 & 0.5163 & 0.4729 & 0.113 & 0.382 & 56.5 \\
    9.98 & 1050 & 0.5166 & 0.4731 & 0.111 & 0.382 & 58.0 \\
    10.00 & 1100 & 0.5169 & 0.4733 & 0.109 & 0.383 & 59.9 \\
    10.02 & 1150 & 0.5172 & 0.4735 & 0.107 & 0.384 & 61.5 \\
    10.05 & 1200 & 0.5175 & 0.4737 & 0.105 & 0.382 & 62.7 \\
    \end{tabular}
    \caption{Seeding results for the critical nucleus size $n^*$, interfacial tension $\gamma$, and barrier height $\beta\Delta G^*$ for melting of hard spheres as a function of pressure $\beta P \sigma_{\mathrm{eff}}^3$. Here $\eta_s$ is the effective packing fraction of the solid phase, $\eta_{fl}$ the effective packing fraction of the fluid phase at the same pressure $\beta P \sigma_{\mathrm{eff}}^3$, and $\beta\Delta\mu_{\mathrm{eff}}$ is the effective driving force for melting under those conditions.}
    \label{tab:seeding}
\end{table}